\documentclass[11pt,a4paper]{article}
\usepackage{jheppub}
\usepackage{amsmath}
\usepackage{amssymb}
\usepackage{xcolor}
\usepackage{float}
\usepackage{enumerate}

\newcommand{\bea}{\begin{eqnarray}}
\newcommand{\eea}{\end{eqnarray}}
\newcommand{\bit}{\begin{itemize}}
\newcommand{\eit}{\end{itemize}}

\def\nl{\nonumber \\}

\def\d{\partial}

\def\a{\alpha}

\def\b{\beta}

\def\s{\sigma}

\def\p{\partial}

\def\li{\mathcal{L}}

\def\le{\left(}
\def\ri{\right)}

\def\beq{\begin{equation}}
\def\eeq{\end{equation}}

\def\arr{{\rightarrow}}

\def\commuA{(\s_{[1} D_A \s_{2]})}

\def\th {\tilde{h}}

\def\E{\eta}
\def\v{\delta}

\title{On Newton-Cartan
local renormalization group \\ and anomalies} 

\author[a,b]{Roberto Auzzi,} 
\author[a]{Stefano Baiguera,} 
\author[a]{Francesco Filippini}
\author[a,c]{and Giuseppe Nardelli}

\affiliation[a]{Dipartimento di Matematica e Fisica, Universit\`a Cattolica
del Sacro Cuore, \\
Via Musei 41, 25121 Brescia, Italy}
\affiliation[b]{INFN Sezione di Perugia, \\ Via A. Pascoli, 06123 Perugia, Italy}
\affiliation[c]{TIFPA - INFN, c/o Dipartimento di Fisica, Universit\`a di Trento, 38123 Povo (TN), Italy}

\emailAdd{roberto.auzzi@unicatt.it}
\emailAdd{giuseppe.nardelli@unicatt.it}
\emailAdd{stefano.baiguera92@gmail.com}
\emailAdd{francesco.filippini@hotmail.it}

\abstract{
Weyl consistency conditions 
are a powerful tool to study the irreversibility properties of the
renormalization group. We apply this formalism to non-relativistic
theories in $2$ spatial dimensions with boost invariance
and dynamical exponent $z=2$. 
Different possibilities are explored, depending on the structure
of the gravitational background used as a source for
the energy-momentum tensor.
}


\begin{document}

\maketitle

\section{Introduction}

Relativistic trace anomalies (see \cite{Duff:1993wm} for a review)
 give non-trivial constraints on the 
possible infrared (IR) dynamics which can emerge from an
ultraviolet (UV) unitary theory. In $2$ dimensions this is established by 
Zamolodchikov $c$-theorem \cite{Zamolodchikov:1986gt}.
In $4$ dimensions the monotonicity property 
of the anomaly coefficient $a$ ($a$-theorem)
was first conjectured  in \cite{Cardy:1988cwa};
a perturbative proof was given by 
\cite{Osborn:1989td,Jack:1990eb,Osborn:1991gm}
with the local renormalization group (RG) equations.
A proof using dispersion relations was given in 
\cite{Komargodski:2011vj,Komargodski:2011xv}.

The local RG equations \cite{Osborn:1989td,Jack:1990eb,Osborn:1991gm}
are derived imposing the Wess-Zumino (WZ) consistency conditions 
for the trace anomaly 
\cite{Bonora:1983ff,Bonora:1985cq}
of the theory in a generic 
gravity background with spacetime-dependent couplings.
It is a very useful tool to study 
relativistic RG flows nearby conformal fixed points
in various dimensions 
\cite{Fortin:2012hn,Luty:2012ww,Jack:2013sha,Baume:2014rla,
Nakayama:2013wda,Stergiou:2016uqq}
and it has also interesting applications
in the supersymmetric case
\cite{Jack:2013sha,Auzzi:2015yia,Gomis:2015yaa}.
For reviews see \cite{Nakayama:2013is,Shore:2016xor}.

Genuine anomalies correspond to terms which are
intrinsic properties of the field theory and do not depend on
the choice of counterterms used in the renormalization procedure.
There are other violations of conformal symmetries in curved space
which instead are counterterm-dependent.
For example, in the relativistic case in $4$ dimensions the trace
 of the energy-momentum
tensor contains the following terms:
\begin{equation}
T^\mu_\mu \supset a E_{4} - c W^{2} + b R^2 -\frac23 a' \, \Box R \, .
\end{equation}
The anomaly coefficient $a$ coincides, at the fixed point, with the decreasing
function in the $a$-theorem, while  $c$ has no monotonicity property \cite{Anselmi:1997am}.
The coefficient $b$ vanishes at conformal 
fixed points due to the Wess-Zumino consistency conditions.
Moreover, as we will show,
 it is possible to choose counterterms in such a way that it vanishes 
along a specific RG flow trajectory which interpolates between an UV 
and an IR fixed point.
The $a'$ coefficient can be shifted by a counterterm 
in the vacuum functional which is proportional to 
$R^2$ and so it is not a genuine scheme-indepedendent anomaly.
However, this term could still play a role in the monotonocity
properties of the RG flow. 
For example, following   \cite{Anselmi:1999xk,Anselmi:2002fk}
one may consider the difference between the UV and IR 
$a'$ coefficient $\Delta a'= a'_{UV}-a'_{IR}$ along a given RG flow trajectory;
in a scheme where $b=0$ along the RG flow,
 this is a monotonically-decreasing quantity.
 Note that it explicitly depends on the given RG trajectory
 and not just on its end points.
In \cite{Anselmi:1999xk,Anselmi:2002fk} 
it was conjectured
that the minimum of $\Delta a'$ among all the possible
RG flows coincides with  $\Delta a=a_{UV}-a_{IR}$ .

It is interesting to explore trace anomalies also in the non-relativistic
case, in order to study non-trivial constraints on the renormalization group flow\footnote{
One may wonder if a non-relativistic trace anomaly
can be obtained as an infrared limit of a relativistic one. This is not the case because the relativistic
scale symmetry is explicitly broken by the 
mass gap which is necessary to have a 
non-relativistic limit.
Indeed relativistic trace anomaly coefficients
count the number of massless degrees of freedom,
 which is zero for a $z \neq 1$ fixed point.
The non-relativistic scale symmetry is an IR emergent phenomenon which does not exist in the far relativistic UV.} 
 At a generic scale-invariant fixed point,
 various  relative scaling of space and time, which are parameterized by the
dynamical exponents $z$,
are in principle allowed i.e.
\beq
x^i \arr e^{\s} x^i \, ,\qquad t \arr e^{z \s} t \, . 
\eeq
Moreover, the details of the anomaly structure depend crucially 
 on the symmetry content of the theories, in particular
if we require or not boost invariance and also if we require or not
integrability of time slices of the gravitational background (Frobenius condition).
This condition is important for causality when the gravitational background is physical;
on the other hand, in order to define the energy-momentum tensor,
we need to consider generic variations of the background metric,
including the ones which do not satisfy the Frobenius condition.

Scale anomalies in theories without boost invariance
(Lifshitz) were studied by several authors, e.g.
\cite{Adam:2009gq, Baggio:2011ha, Griffin:2011xs, Arav:2014goa, Arav:2016xjc, Pal:2016rpz}.
It turns out that, in all the cases that have been studied so far,
 the scheme-independent trace anomalies
at the fixed point   have vanishing Weyl variation (type $B$  \cite{Deser:1993yx}\footnote{Anomalies with non-trivial Weyl variation are
instead called of type $A$; they correspond to non-trivial solutions
of the Wess-Zumino consistency conditions.}).

In this paper we will be interested in the case with boost invariance.
The natural background is provided by the Newton-Cartan (NC) gravity.
Two different attitudes concerning trace anomalies
are possible:
\begin{itemize}
\item The case where causality is not required 
on the gravity background;
 this setting was first studied by \cite{Jensen:2014hqa}.
 In this case the anomaly in $2+1$ dimensions and for $z=2$
  has a very rich  structure, because
 an infinite number of terms can be written by dimensional analysis.
 All these terms live in separated sectors, 
 which are labelled by the integer $N_n$.
A Weyl variation does not change the value of $N_n$
 and the WZ consistency conditions can be studied independently
 in each separated sector, where just a finite number of terms is present.
  In particular, in the simplest
 sector $N_n=0$  the anomaly structure is identical
 to the trace anomaly of relativistic theories in $4$ dimensions
 and a natural type $A$
 candidate for a monotonicity theorem is the coefficient
 of the $E_4$ term.  
 The calculation of the $N_n=0$ anomaly in the case of a free 
 scalar was recently done in \cite{Auzzi:2016lxb}.
\item The case in which the Frobenius condition is imposed
has a much simpler anomaly structure. For $d=2+1$ and $z=2$,
 the number of terms allowed by dimensional analysis is finite \cite{Auzzi:2015fgg}
and the only scheme-independent anomaly turns out to be of type $B$
\cite{Arav:2016xjc,Auzzi:2015fgg}. 
\end{itemize}

The purpose of the present paper is to initiate an analysis, 
using the local RG formalism, of the non-relativistic scale anomalies
in theories with Galilean boost invariance.
We focus on the $d=2+1$ and $z=2$ case, and we explore both
the case with and without Frobenius condition:
\begin{itemize}
\item
The structure of the local RG equation in the simplest
sector $N_n=0$
of the case without Frobenius conditions turns out
to be the same as for the relativistic theories in $4$ dimensions. 
\item
If we impose the Frobenius conditions, 
there is no scheme-independent $a$-theorem candidate.
However it is still interesting to study the local RG equations;
for example it might be possible to identify
monotonic scheme-dependent quantities analog
to the $a'$ of the relativistic four-dimensional case.
A similar study, in the case without boost invariance,
was done in \cite{Pal:2016rpz}.
\end{itemize}

We also study in detail the $N_n=1$ sector of the anomaly 
 without Frobenius conditions and we find that no anomaly is allowed
 by WZ consistency condition. We leave the study
 of higher sectors with $N_n>1$ as a challenging problem for further
 investigation; it could be that these sectors contain some interesting 
  candidates for monotonicity theorems.

\section{Preliminaries}

\subsection{Newton-Cartan geometry}

The Newton-Cartan (NC) gravity is a covariant version of Newtonian 
gravity; putting non-relativistic theories in a Newton-Cartan
gravitational background is a very useful
tool in condensed-matter physics because it gives the natural sources
for the operators in the energy-momentum tensor multiplet, e.g.
 \cite{Son:2005rv,Hoyos:2011ez,Son:2013rqa,Geracie:2014nka,Jensen:2014aia,
 Hartong:2014pma,Hartong:2014oma,Hartong:2015wxa}. 

A NC gravity background is defined by 
the tensors $h_{\a \b}$, $h^{\mu \nu}$,
and by the vectors $v^\mu$ and $n_\mu$ with properties:
\beq 
n_\mu h^{\mu \a}=0 \, ,
\qquad
n_\mu v^\mu =1 \, ,
\label{orto}
\eeq
\beq
h^{\mu \a}h_{\a \nu}=\delta^\mu_\nu -v^\mu n_\nu=P^\mu_\nu \, , \qquad
h_{\mu \a} v^\a =0 \, .
 \eeq
  As a further ingredient,
  a $U(1)$ gauge potential $A_\mu$ for the 
  number particle symmetry must be introduced.
Causality is specified by the Frobenius condition, which takes the form
\beq
dn \wedge n=0 \, ,
\label{frob}
\eeq
 where the $1$-form $n=n_\mu dx^\mu$  specifies
the time direction.

 The symmetries of NC gravity include a local version of Galilean boosts,
 which are called Milne boosts and are given by the following shifts, parameterized
 by the functions $\psi_\mu(x^\a)$:
 \bea
v'^\mu & = & v^\mu+h^{\mu \nu} \psi_\nu \, \nl
h'_{\mu \nu} & = & h_{\mu \nu} -(n_\mu P_\nu^\rho+ n_\nu P_\mu^\rho) \psi_\rho
+n_\mu n_\nu h^{\rho \sigma} \psi_\rho \psi_\sigma \, , \nl
A'_\mu & = & A_\mu+P^\rho_\mu \psi_\rho -\frac{1}{2} n_\mu h^{\a \b} \psi_\a \psi_\b \, . 
\eea
As a tool to write Milne boost invariant quantities in a systematic 
way, we use the null reduction from a relativistic parent
space\footnote{Capital latin indices will always refer to $d+2$-dimensional
tensors, while greek ones to $d+1$-dimensional objects.}, 
as introduced in \cite{Duval:1984cj}:
\beq
\label{DLCQ01}
G_{MN}  = \left(
\begin{array}{cc}
 0& n_\mu \\ n_\nu \ \  & n_\mu A_\nu + n_\nu A_\mu + h_{\mu \nu} \\
  \end{array}\right) \, , \qquad
  G^{MN}  = 
  \left(
\begin{array}{cc}
 A^2-2 v \cdot A \ \ & v^\mu - h^{\mu \sigma} A_\sigma \\ 
v^\nu - h^{\nu \sigma} A_\sigma  &  h^{\mu \nu} \\
  \end{array}\right)  \, ,
\eeq
where the null reduction is taken along the direction:
\beq
n^M=(1,0, \dots) \, , \qquad n_{M}=(0,n_\mu) \, 
\eeq
We will refer to this trick as 
Discrete Light-Cone Quantization (DLCQ).
Let $D_A$, $R$, $R_{ABCD}$, $R_{AB}$ denote respectively  the covariant derivative,
the scalar curvature and the Riemann and Ricci tensors  defined by the 
Levi-Civita connection from the metric in eq.~(\ref{DLCQ01}).
The spacetime volume element
is defined as:
\beq
\sqrt{g}=\sqrt{\det (n_\mu n_\nu + h_{\mu \nu})} =\sqrt{-\det G_{AB}}\, .
\eeq

Local Weyl transformations are parameterized by
a function $\s$.
In the DLCQ formalism $\s$ is
 taken independent from the null direction:
\beq
n^A D_A \s =0 \, .
\eeq
In our conventions, the Weyl  scaling of the extra-dimensional metric 
and of the NC objects is:
\beq
G_{MN} \arr e^{2 \s} G_{MN} \, , \qquad
n_\mu \arr e^{2 \s} n_\mu \, , \qquad
h_{\mu \nu} \arr e^{2 \s} h_{\mu \nu}  \, .
\eeq
Here and in the rest of the paper we specialize to dynamical exponent $z=2$.

\subsection{Sources}

The allowed perturbations on the background fields must satisfy
eqs.~(\ref{orto}). The most general  
variation is parameterized by
 an arbitrary $\delta n_\mu$, a transverse
 perturbation $\delta u^\mu$ with
 $\delta u^\mu n_\mu =0$ and a transverse metric 
 perturbation $\delta \th^{\a \b} n_\b=0$.
 In term of these quantities, the variations of the background fields read:
 \beq
 \delta A_\mu \, , \qquad
 \delta n_\mu \, , \qquad
 \delta v^\mu = - v^\mu v^\a \delta n_\a + \delta u^\mu \, , \qquad
 \delta h^{\mu \nu} =-v^{\mu} \delta n^{\nu}-\delta n^\mu v^\nu 
 -\delta \th^{\mu \nu} \, .
 \label{varii}
 \eeq 
 At the linear order nearby the flat limit, eq.~(\ref{varii}) 
 gives\footnote{Upper and lower spatial indices $i,j$ are raised by Kronecker delta and so are interchangeable.}:
 \bea
 n_\mu  &=&  (1+\delta n_0 , \delta n_i) \, , \qquad
 v^\mu =(1-\delta n_0, \delta u_i) \, , \qquad \delta \th^{0 i}=0 \, ,
 \nl
  h_{\mu \nu} &=&
  \left(
\begin{array}{cc}
 0& -\delta u_i \\ -\delta u_i  &  \delta_{ij} + \delta \th_{ij} \\
  \end{array}\right) \, , \qquad
h^{\mu \nu}=
  \left(
\begin{array}{cc}
 0& -\delta n_i \\ -\delta n_i  &  \delta_{ij} - \delta \th_{ij} \\
  \end{array}\right) \, .
\eea 
In term of the null reduction fields, this corresponds to:
 \bea
G_{A B} & = &
  \left(
\begin{array}{ccc}
0 & 1 + \delta n_0 &  \delta n_i \\
1+\delta n_0 & 2 \delta A_0 & \delta A_i -\delta u_i \\ 
\delta n_i  & \delta A_i -\delta u_i &\delta_{ij} + \delta \th_{ij} \\
  \end{array}\right) \, ,
  \nl
  G^{A B} & = &
  \left(
\begin{array}{ccc}
-2 \delta A_0 & 1- \delta n_0 & -\delta A_i+ \delta u_i \\
1-\delta n_0 &  0& -\delta n_i \\ 
-\delta A_i+ \delta u_i  &  -\delta n_i&\delta_{ij} + \delta \th_{ij} \\
  \end{array}\right) \, .
  \label{perturba}
\eea

These sources are useful to define conserved currents.
We will consider the vacuum functional $W[g_{\mu \nu}]$:
\beq
e^{i W[G_{MN}]}= \int {\cal D} \phi \,  e^{i S[\phi, G_{M N}]}
\eeq
where $\phi$ runs over the dynamical fields of the theory.
The expectation values of the
energy-momentum tensor multiplet are defined by: 
\beq
\delta W = \int d^{d+1} x\sqrt{g} \le
\frac{1}{2} T_{i j} \delta \th_{i j} + j^\mu \delta A_\mu
-\epsilon^\mu \delta n_\mu - p_i \delta u_i 
\ri \, .
\eeq
In this expression $p_i$ is the momentum density,
$T_{i j}$ is the spatial stress tensor, $j^\mu=(j^0,j^i)$ contains
the number density and current and $\epsilon^\mu=(\epsilon^0,\epsilon^i)$
 the energy density and current. $j^i$ is proportional to 
$p_i$  because only the combination
 $\delta A_i - \delta u_i$ enters the DLCQ metric,
 see eq.~(\ref{perturba}).

The first-order Weyl variation $\Delta$
of the vacuum functional nearby flat spacetime is:
 \beq
\Delta_W W= 2 \s G_{AB} \frac{\delta W}{\delta G_{AB}} = 
2 \s \le \delta^{ij} \frac{\delta W}{\delta (\delta \th_{ij}) } +
2 \frac{\delta W}{\delta (\delta n_0)} \ri =
2 \s( T^i_i -2 \epsilon^0) \, .
\eeq
In the rest of the paper we specialize to the $d=2$ case.

\subsection{The trace anomaly without Frobenius condition}

\label{anoma1}

Let us consider a conformal field theory coupled to a
background NC geometry. We will restrict to the parity-invariant case.
A generic term inside the anomaly in $d+1$ dimensions
can be written as a scalar 
obtained contracting the following $d+2$-dimensional tensors:
the curvature $R_{ABCD}$, the null direction $n_A$
and the metric $G^{AB}$.
We denote respectively by $ N_n $,  $ N_D $ and $ N_R $ 
the numbers  of $ n_A $ vectors,
 covariant derivatives and  Riemann tensors\footnote{In the notation of \cite{Arav:2016xjc}:
  $N_{T}=-N_{n}$,
  $N_{S}= N_{D} +2 N_{R} + N_{n}$,
   $N_\epsilon=0$.}, all taken with lower indices.
 $N_G$ denotes the number of metric tensors
 (all taken with upper indices) which are used for the contraction.
The condition for a term to be a scalar is
\begin{equation}
4N_{R} +N_{n} +N_{D} = 2 N_{G}  \, ,
\label{Matching of indices}
\end{equation}
while the requirement of having the correct Weyl weight 
in order to enter the anomaly is given by
\begin{equation}
2 N_{R} +2 N_{n}-2 N_{G} = -4  \, .
\label{Matching of Weyl weight}
\end{equation}
Eliminating $N_G$ from 
eqs.~(\ref{Matching of indices},\ref{Matching of Weyl weight}),
one obtains
\beq
4=N_D+2 N_R -N_n \, .
\label{constr}
\eeq

The number $N_n$ is unchanged by Weyl transformation.
This observation has interesting implications: the 
Wess-Zumino consistency conditions of the terms with different
$N_n$ do not mix with each other.
Consequently, we can study each sector with a different value
of $N_n$ independently.

The basis for the anomaly in the sector $ N_n=0 $ and the cohomological problem are exactly the same as the relativistic trace anomaly in (3+1)-dimensional space-time \cite{Jensen:2014hqa}: 
\begin{equation}
\mathcal{A} = a E_{4} - c W^{2} + \mathcal{A}_{\mathrm{ct}} \, ,
\qquad
\mathcal{A}_{\mathrm{ct}}=-\frac23 a' \, \Box R \, \, ,
\label{anoma0}
\end{equation}
where the Euler density and the Weyl tensor are calculated 
in the DLCQ space-time starting from the corresponding metric
eq.~(\ref{DLCQ01}):
\beq
E_4=R^2_{ABMN} -4 R^2_{AB} +R^2 \, ,
\qquad
W^2=W^2_{ABMN}=R^2_{ABMN} -2 R^2_{AB} +\frac{1}{3} R^2 \, .
\eeq
 In this way we immediately recognize the existence of a type A anomaly, 
 the $ E_4 $ term, and of a type B anomaly, the squared Weyl tensor.
The anomaly in eq.~(\ref{anoma0}) was explicitly computed for a free
scalar in \cite{Auzzi:2016lxb}.

The sectors with $N_n>0$ will be discussed in section \ref{greaterNn}.

\subsection{The trace anomaly with Frobenius condition}

\label{anoma2}

If we impose the causality condition (\ref{frob})
the structure of the anomaly drastically changes:
only a finite number of non-vanishing  terms are allowed
by the conformal dimension. Moreover the type $A$
anomaly can be  eliminated by a local counterterm and becomes
scheme-dependent. This case was studied in \cite{Auzzi:2015fgg}
and in \cite{Arav:2016xjc}  with different formalisms.
In this section we will summarize the results using the notation
of \cite{Auzzi:2015fgg}.

The condition that $n_M$
is a Killing vector for the metric gives:
\beq
0=\li_n (G_{M N})=D_M n_N+D_N n_M \, .
\eeq
The condition that $n_A$ is null gives
$D_M n^M=0$ and $n^S D_S n_M=0$.
We define the 2-form 
\beq
\tilde{F}=\tilde{F}_{AB} dx^A \wedge dx^B =2 d n \, ,
\eeq
which in components reads:
\beq
\tilde{F}_{MN}=\p_M n_N -\p_N n_M=2 D_M n_N \,  .
\label{FTILDE}
\eeq
In particular,  $\tilde{F}_{- \a} =0$. 

The causality condition $n \wedge dn=0$
implies that $dn=n \wedge w$
for some one form $w$:
\beq 
\tilde{F}_{A B} = n_{[A} w_{B]} \, , \qquad w_A=(0, w_\a) \, ,
\qquad n^A w_A = 0 \, .
 \label{semp0}
\eeq
Note that $w_A$ in eq.~(\ref{semp0}) is not uniquely determined;
for example it could be shifted by
\beq
w_A \rightarrow w_A+p n_A \, ,
\label{shift}
\eeq
 where $p$ is an arbitrary function 
 ($x^-$ independent)
without affecting $\tilde{F}_{AB}$.

We define:
\beq
 \chi=\frac{1}{16} G^{MN} w_M w_N\, .
 \label{semp1}
\eeq
\beq
\Omega_{A B}=\frac{1}{16} (w_A w_B -4 D_A w_B) \, ,  \qquad
\Omega=\Omega_{AB} G^{AB} \, .
\label{omiga}
\eeq
Note that $\chi$ is invariant under the shift
in eq.~(\ref{shift}).
We can use the ambiguity in eq.~(\ref{shift})
to render $\Omega_{AB}$ symmetric.
The following property is useful:
\beq
\Omega_{AB} n^B =\Omega_{B A} n^B= \chi \, n_A \, ,
\label{omiga2}
\eeq
It is convenient to introduce 
\beq
 J  =\Omega-2 \chi +\frac{R}{6}\, ,
  \label{jj}
 \eeq
whose Weyl variation is $-2 \s J$.

By dimensional analysis, the anomaly
can be written as a finite linear combination:
\beq
\mathcal{A}_\s =\sum_{k=1}^{12} b_k \mathcal{A}_\s^k \, ,
\qquad
 \mathcal{A}_\s^k =\int \sqrt{g}  \, d^{3}x \, \left(\s \, A_k \right) \, ,
\eeq
where $b_k$ are the anomaly coefficients and 
\bea
 &  &  A_1=D^2 R \, , 
\qquad   A_2=R^2  \, , \qquad 
A_3=\chi^2 \, ,   \nl
& & A_4=\Omega^2 \, , \qquad  A_5=\chi \Omega  \, , \qquad
A_6=\chi R \, , \nl
& & A_7=\Omega R \, , \qquad
A_{8}=\Omega_{AB} \Omega^{AB} \, ,  \qquad
A_{9}= \Omega_{AB} w^A w^B \, ,
  \nl  
& & 
 A_{10} = w^A D_A R \, ,  \qquad  
A_{11}=D^2 \chi \, , \qquad
A_{12}=D^2 \Omega \, . 
\label{BAE}
\eea
In order to determine this basis,
it is important to use the following
relations, which are valid only if
the Frobenius condition is satisfied:
\bea
&& W^2=12 J^2 \, ,
\nl
&& E_4=72 \chi^2 -4 \chi R -48 \chi \Omega + 8 \Omega^2 -8 \Omega_{AB} \Omega^{AB} \, ,
\nl
&& (R_{AB} +2 \Omega_{AB}) w^A w^B = 8 \chi(R-6 \chi +4 \Omega) \, ,
\nl
&& \Omega_{AB} (R^{AB} +2 \Omega^{AB})
= 12 \chi^2 +\frac{1}{2} \Omega (R+4 \Omega) - \chi (R + 9 \Omega) \, .
\eea

Imposing the Wess-Zumino consistency condition,
we find that at the conformal fixed point 
the anomaly is of type B:
\beq
\mathcal{A}= b  \s J^2  + \mathcal{A}_{\rm ct}
\eeq
where the terms in $\mathcal{A}_{\rm ct}$ 
 are arbitrary linear combinations of
 \bea
\label{schemedependent} 
&& \s D^2 R \, , \qquad \s D^2 (\Omega-2 \chi) \, , \qquad
\s \le 12 \chi^2-4 \chi \Omega-\frac{1}{2} \Omega_{A B} w^A w^B \ri\, ,
\nl
&& \s \le 2 R \chi -2 R \Omega +\frac{w^A D_A R}{2} -6 D^2 \chi \ri \, , \nl
&&
\s \le-9 \chi^2 -\Omega^2+6 \chi \Omega+\Omega_{AB}^2 
+\frac{ \chi R}{2} \ri
 \, ,
\eea
which can all be written as Weyl variation of
local counterterms
(see appendix \ref{controtermini}).

\section{Local RG without Frobenius condition and $N_n=0$}

The couplings $g^i$ are now taken as space-time dependent sources for the 
marginal operators $\mathcal{O}_i$ of the theory, e.g.:
\beq
\langle \mathcal{O}_i (x) \rangle = \frac{\delta W}{\delta g^i(x)} \, ,
\eeq
where $W$ is the vacuum functional.
The local RG generator is:
\beq
-\int \sqrt{g} d^3 x \le 
\Delta_W W +  \Delta_\b W
 \ri = \mathcal{A}
\eeq
where
\beq
\Delta_W W= 2 \s G_{AB} \frac{\delta W}{\delta G_{AB}} \, , \qquad
\Delta_\b W=\s \b^k
\frac{\delta W}{\delta g_k} \, , 
\eeq
and $\beta^k$ denote the beta functions of each coupling: $\beta^k=\frac{d g^k}{d \log \mu}$.
The anomaly $ \mathcal{A}$ now includes also terms with space-time
derivatives of the couplings $g^k$.
To avoid confusion, starting from this section, the lowercase
latin indices $i,j,k, \dots$ run on the space of the couplings $g^i$.
Moreover, we denote by $\p_i f=\frac{\p f}{\p g^i}$
the derivatives of functions with respect to couplings.
If the coupling $g^i$ has some charge $m$
under the "mass" particle number,
 a dependence $g^i=\tilde{g}^i(x^\mu) e^{i m x^-}$
 should be given in the null direction.
 Note that for $m=0$ the coupling $g^i$ can be real, while
 for $m \neq 0$ it must be complex.

In the sector $N_n=0$,
the anomaly part of the local RG equation 
can be formally obtained by DLCQ reduction
of the relativistic local RG \cite{Osborn:1991gm}:
\beq
\mathcal{A}= \mathcal{B} + (D_A \sigma) \mathcal{Z}^A +(D^2 \s) \mathcal{X} \, ,
\eeq
where
\bea
&\mathcal{B}&=  a E_{4} - c W_{ABCD}^{2}
+\frac{b}{9} R^2 + \frac{1}{3} {\chi^e}_i  D_A g^i D^A R
+\frac{1}{6} \chi^f_{ij} D_A g^i D^A g^j R
+\frac{1}{2} \chi^g_{ij} D_A g^i D^A g^j
\nl
& &
+\frac{1}{2} \chi^a_{ij} D^2 g^i D^2 g^j 
+\frac{1}{2} \chi_{ijk}^b D_A g^i D^A g^j D^2 g^k 
+\frac{1}{4} \chi^c_{ijkl} D_A g^i D^A g^j D_B g^k D^B g^l \, ,
\nl
&\mathcal{Z}^A& = w_i E^{AB} D_B g^i +\frac{1}{3} Y_i R D^A g^i
+S_{ij} D^A g^i D^2 g^j +\frac{1}{2} T_{ijk} D^A g^i D_B g^j D^B g^k \, ,
\nl
&\mathcal{X}&=-\frac{2 a'}{3} R - U_i D^2 g^i -\frac{1}{2} V_{ij} D_B g^i d^B g^j \, .
\label{loca-osb}
\eea

Formally we can write exactly the same consistency conditions
as in the relativistic case in $d=4$, and the same set of local
counterterms can be used to shift the anomaly coefficients.
A detailed analysis can be found in \cite{Osborn:1991gm}.
It is interesting to focus on the consistency conditions that give
a perturbative proof of the $a$ theorem:
\beq
8 \p_i a-\chi^g_{ij} \b^j =-\li_\beta w_i \, ,  
\label{a-th-1}
\eeq
\beq
\chi^g_{ij} +2 \chi^a_{ij} +2 \p_i \b^k \chi^a_{kj} + \b^k \chi^b_{kij} = \li_\beta  S_{ij} \, ,
\label{a-th-2}
\eeq
where $\li_\beta$ is the Lie derivative in the coupling space 
along the direction given by the beta functions $\beta^i$, e.g.:
\beq 
\li_\b w_i = \b^k \p_k w_i + \p_i \b^k w_k \, .
\eeq
Then equation (\ref{a-th-1}) can be re-written as:
\beq
\b^i \p_i \tilde{a}=\frac{1}{8} \chi^g_{ij} \b^i \b^j \, ,
\qquad
\tilde{a}=a+\frac{1}{8} w_i \b^i \, ,
\eeq
then the quantity $ \tilde{a}$ is monotonically decreasing along the RG flow
provided that $ \chi^g_{ij}$ is a positive-definite metric.
In the limit of small $\beta^i$ (which corresponds to leading-order in conformal 
perturbation theory), this is equivalent to the fact that $ \chi^a_{ij}$
is negative-definite (see eq.~\ref{a-th-2}). 

In the relativistic case, the positiveness of $-\chi^a_{ij}$ 
follows from the fact that it coincides, 
 with the Zamolodchikov metric\footnote{The Zamolodchikov metric is defined
from the two-point functions of 
the marginal operators: $ \langle \mathcal{O}_i(x) \mathcal{O}_i(0) \rangle =
\frac{G_{ij}(g^k)}{x^{2 (d+2)}}$. } in an opportune class of schemes.
Positivity of Zamolodchikov metric,
for unitary theories, in turn
follows from the positivity of the spectral density function
in the K\"all\'en-Lehmann spectral representation, see \cite{Shore:2016xor} for a review.

The key missing ingredient for a perturbative proof of the
monotonicity of $\tilde{a}$ in the non-relativistic case
is the negativity of the anomaly coefficient $ \chi^a_{ij}$.
In spite of the apparent similarities with the relativistic case,
the proof of the negativity of $\chi^a_{ij}$
does not seem to follow from a straightforward generalization
of the relativistic argument.
We leave this issue as a topic for further investigation.

Another interesting consistency condition is:
\beq
\b^k \p_k (4 a' + U_i \b^i)=8 b -\chi^a_{ij} \b^i \b^j \, .
\label{inte}
\eeq
The quantity $a'$ is scheme-dependent at the fixed point;
indeed if we add a counterterm 
\beq 
\delta W = C R^2+\frac13 E_i D_A g^i  D^A R \, ,
\eeq
there is the following shift in the anomaly coefficients in eq.~(\ref{loca-osb}):
\beq
\delta d= 2 C +\frac12 E_i \b^i \, , \qquad \delta b = \li_\b(D) \,  ,
\qquad \delta  \chi^e_i =\li_\b (E_i) \, .
\eeq
The quantity $b$ vanishes at the conformal fixed point;
then it can be written as $b=\b^i \xi_i$ for some
smooth vector $\xi_i$. Along a particular RG trajectory,
we can write $\xi_i=\p_i G$, and then we can use the counterterm
$C=-G$ to make $b$ vanishing along an RG trajectory.
In this scheme the difference $\Delta a'=a'_{UV}-a'_{IR}$
corresponds to  
\beq
\Delta a'= -\frac14 \int_{IR}^{UV}  \chi^a_{ij} \frac{d g^i }{dt} \frac{d g^j}{d t} \, dt \, , \qquad t=\log \mu \, ,
\eeq
which formally resembles the action of a free particle moving in the space of couplings.
If the metric $-\chi^a_{ij}$ is positive-definite,
then $\Delta a'$ is always positive, even if dependent on the specific 
RG flow trajectory.
In the relativistic case, it was conjectured  \cite{Anselmi:1999xk,Anselmi:2002fk}
that $\Delta a \leq \Delta a'$ and that the lowest possible value of $\Delta a'$
coincides with $\Delta a$; it is reasonable that a similar property is valid
also in the non-relativistic case.

\section{Local RG with Frobenius condition}

If the Frobenius condition holds, the number of terms in the anomaly is finite.
In this section we perform an analysis of the local renormalization group
in the case of marginal perturbations; in this case the study of the WZ consistency
conditions is formally different from the relativistic case.
We know from section \ref{anoma2} that the scheme-independent content of the anomaly is just a single term of type $B$, 
whose coefficient is therefore unconstrained by the local RG equations. 
On the other hand, $T^\mu_\mu$ has several geometrical contributions
which have a non-trivial RG evolution, analogously
 to the $a'$ coefficients of the relativistic case.
These coefficients are not universal properties of 
each conformal fixed point, but still may give interesting
constraint on the possible RG flows.

We choose the following basis:
\begin{equation}
(\Delta_{\s}^{W} +\Delta_{\s}^{\beta}) W =    \int d^{3} x \sqrt{g} \, 
\le
\s \mathcal{B} \cdot \mathcal{R} + \, D_{A} \s \mathcal{Z}^{A} 
\ri \, ,
\label{LRG-frobe}
\end{equation}
where
\bea
\mathcal{B} \cdot \mathcal{R} &=& 
  \E^1 R^{2} + \E^2 \chi^{2}  + \E^3 \Omega^{2} + \E^4 \chi \Omega + \E^5 \chi R + \E^6 \Omega R 
  \nl
 && + \E^7 \Omega_{AB} \Omega^{AB} + \E^8 \Omega_{AB} w^{A} w^{B} + \E^9  w^{A} D_{A} R 
 + \E^{10} D^2 R + \E^{11} D^2 \chi + \E^{12} D^2 \Omega 
 \nl
  && + c_{i}^{1} D^{2} g^{i} R  + c_{i}^{2} D^{2} g^{i} \chi +
  c_{i}^{3} D^{2} g^{i} \Omega  + c_{i}^{4} D_{A} g^{i} w^{A} R 
  + c_{i}^{5} D_{A} g^{i} w^{A} \chi 
   + c_{i}^{6} D_{A} g^{i} w^{A} \Omega  
   \nl
  && +e_{ij}^{1} D_{A} g^{i} D^{A}g^{j} R  +  e_{ij}^{2} D_{A} g^{i} D^{A}g^{j} \chi + e_{ij}^{3} D_{A} g^{i} D^{A}g^{j} \Omega   
 \nl
  && + e_{ij}^{4} D_{A} g^{i} D_{B}g^{j} w^{A} w^{B}  +
  e_{ij}^{5} D_{A} g^{i} D_{B}g^{j} \Omega^{AB} + e_{ij}^{6} D^{2} g^{i} D^{2} g^{j} 
  +e_{ij}^{7} D^{2} g^{i} D^{A} g^{j} w_{A} 
  \nl
   && + k^1_{ijm} D_{A}g^{i} D^{A}g^{j} D^{2} g^{m} 
  + k^2_{ijm} D_{A}g^{i} D^{A}g^{j} D^{B} g^{m} w_{B} 
  \nl
  && + q_{ijmn} D_{A}g^{i} D^{A}g^{j} D_{B} g^{m} D^{B}g^{n} \, ,
\eea
\bea
\mathcal{Z}^{A} &=& 
 d_{i}^{1} D_{B}g^{i} \Omega^{AB}  + d_{i}^{2} D_{B}g^{i} w^{A} w^{B}+
d_{i}^{3} D^{A}g^{i} R + d_{i}^{4} D^{A}g^{i} \chi \nl
&& + d_{i}^{5} D^{A}g^{i} \Omega  + D^{A} (d_{i}^{6} D^{2} g^{i})  + d_{i}^{7} D^{2}g^{i} w^{A} 
 \nl
&& + f_{ij}^{1} D^{A}g^{i} D^{2} g^{j}  + D^{A} (f_{ij}^{2} D_{B} g^{i} D^{B}g^{j}) +
f_{ij}^{3} D^{A}g^{i} D^{B} g^{j} w_{B}  
\nl
&& + T_{ijk} D_{B}g^{i} D^{B} g^{j} D^{A} g^{k} \, .
\eea

The terms $\E^k$ are present also in the case
of constant space-time couplings and corresponds
to the anomaly discussed in section
\ref{anoma2}.
At the conformal fixed point, 
all the allowed 
terms with non-vanishing Weyl variations
are variations of local counterterms,
see appendix \ref{controtermini}.

Part of the WZ consistency condition give some algebraic
relations which can be used to solve for
$\E^k$, $k=4,\dots 9$:
\bea
\E^4 &=& \frac{1}{6} (-6 \b^i c_i^1
-  \b^i c_i^3 +2
   \b^i c_i^5 +8 \b^i c_i^6+72 \E^1-2 \E^2-18
   \E^3) \, , \nl  
\E^5 &=& \frac{1}{12} (6 \b^i c_i^1+2
  \b^i c_i^2 + \b^i c_i^3+4 \b^i c_i^6-8 \b^i d_i^7
   -72 \E^1-4
   \E^{11}-8 \E^{12}-6 \E^3) \, , \nl
   \E^6 & = &
   \frac{1}{3} (-3 \b^i c_i^1
  +2 \b^i d_i^7
 +36 \E^1+\E^{11}+2 \E^{12}) \, , \nl 
   \E^7 & = & \frac{1}{2} (-6
   \b^i c_i^1
   -\b^i c_i^3 +72 \E^1-2 \E^3) \, , \nl
   \E^8 & = &  \frac{1}{48} (30 \b^i c_i^1
   +5   \b^i c_i^3+2 \b^i c_i^5 -4
   \b^i c_i^6 -360 \E^1-2 \E^2 +18 \E^3) \, , \nl
   \E^9 &=&  \frac{1}{12} (-2 \b^i d_i^7 -\E^{11}-2
   \E^{12}) \, .
   \label{etaeta}
\eea

The remaining consistency conditions are listed in
appendix \ref{full-list}.
Starting from these, we can identify 
some possible candidates for monotonicity properties.
Let us introduce:
\bea
\tilde{A}_1 &=& -2 \E^{12} +12 \E^{10}-2 \b^k d_k^6 \, , \nl
\tilde{A}_2 &=&
48  \E^1 +\frac{2}{3} \E^2 -4 \E^3 +
\b^k \le -4 c_k^1  -\frac{2}{3} c_k^3 -\frac{2}{3} c_k^5
 +\frac{4}{3} c_k^6  -d_k^1  -16 d_k^2 \ri\, , \nl
\tilde{A}_3 &=&
-72 \E^1  + 2 \E^3
  -4 \E^{11} -8 \E^{12} +\b^k( 6 c_k^1 + c_k^3 
  -d_k^1 -8 d_k^7) \, , \nl
  \tilde{A}_4 &=&
   \E^{11}+2 \E^{12} -6 \E^{10}
+ \b^k(3 d_k^3 -3  c_k^1
  +2  d_k^7) \, ,
\eea
From eq.~(\ref{bo4-bis}), (\ref{sempli}),  (\ref{bo8-bis}),  (\ref{bo1-bis})
we find the following relations:
 \bea
  \b^i \d_i \tilde{A}_1   
 + \b^i( 12 c_i^1 - 2 c_i^3 )
  -4 e_{ij}^6 \b^i \beta^j &=& 0   \, , \nl
\b^i \d_i \tilde{A}_2
  + (32 e_{ij}^4 +2 e_{ij}^5 ) \beta^j \b^i 
 -  (8  c_i^5  + 16 c_i^6) \b^i &=& 0  \, , \nl 
 \b^i \d_i \tilde{A}_3
 +\b^i( - 2 c_i^2 -4 c_i^3 
 -24 c_i^4 +4c_i^6)  
  + (2 e_{ij}^5 
 +8 e_{(ij)}^7 ) \b^i \beta^j &=& 0 \, , \nl 
 \b^i \p_i \tilde{A}_4
+\b^i(12 c_i^4  -6 c_i^1 
- 3 c_l^1 (\d_i \beta^l) 
+3  (\d_i c_l^1) \beta^l) 
 - 6 e_{ij}^1 \b^i \beta^j &=&0   \, ,
\eea
Each of the previous equations has a structure which
is reminiscent of the relativistic equation for $a'$,
see eq.~(\ref{inte}). Note in fact that the terms proportional 
to $c_i^k$ can be shifted using the counterterms in appendix
\ref{controtermini} and set to zero along a RG trajectory.
Consequently, we may consider  linear combinations 
of the four previous quantities $\tilde{A}_k$
as candidates for monotonicity theorems.
In particular, $e^6_{ij}$ is the analog of 
the $\frac12 \chi^a_{ij}$ of the previous section
because it is the coefficient of the $\s D^2 g^i D^2 g^j$ term of the anomaly,
so $\tilde{A}_1 $ is identified with the $a'$ term.


\section{The $N_n> 0$ sectors}

\label{greaterNn}

In the absence of Frobenius condition (\ref{frob}), the number of possible  terms  in the anomaly basis is infinite.
Each term can be formed solely by a combination of vectors $n_A$, 
covariant derivatives $D_A$, curvature tensors $R_{ABCD}$ and metric tensors $G^{AB}$.
These terms can be conveniently organized and classified.
A convenient way is to introduce an integer number $N_n$ labelling the number of DLCQ 
vectors $n_A$ entering the term. 
 This partition splits the infinite terms into 
infinite sectors identified by the value of $N_n$. This has a double advantage: a) 
Since a Weyl transformation does not modify the value of $N_n$, anomalies can be studied separately in each sector $N_n$; b) Although the number of sectors is infinite, the possible anomaly terms are {\it finite} within a given sector.

Inside each sector, the following relations limit the number of possible terms:
\begin{enumerate}[I]
\item the internal symmetries of the curvature tensors,
\item the Bianchi identity 
\begin{equation}
D_E R_{ABCD} + D_C R_{ABDE} + D_D R_{ABEC} = 0 \, ,
\label{Bianchi}
\end{equation}
\item since $n^A$ generates an isometry, the following Lie derivatives vanish
\begin{equation}
\begin{aligned}
& \mathcal{L}_{n}(G_{AB})= 0 \, ,  \\
& \mathcal{L}_{n}(R)= 0 \, ,  \\
& \mathcal{L}_{n}(R_{AB})=0 \, ,  \\
& \mathcal{L}_{n}(R_{ABCD})= 0 \, .   \\
\end{aligned}
\label{Tutte le derivate di Lie}
\end{equation}
\end{enumerate}

Let us fix a value of $N_n$. Then, a dimensional analysis constraints the number $N_R$ of curvature tensors entering the anomaly term to the following
\begin{equation}
\label{Teorema sottosettori}
0 \leq N_R \leq \begin{cases}
\frac{1}{2} (N_n +3)   & \mathrm{if} \, \, N_n \,\, \mathrm{odd} \\
\frac{1}{2} (N_n +4)   & \mathrm{if} \, \, N_n \,\, \mathrm{even} \\
\end{cases}
\end{equation}

This splits the sector $N_n$ into distinct subsectors labelled by $N_R$.
Once  $N_R$ is fixed, the number of covariant derivatives $N_D$ is also fixed according to eq. (\ref{constr}).
We will often make use of the following commutation relation to move terms into different subsectors $N_R$
\begin{equation}
\begin{aligned}
\left[ D_{M},D_{N} \right] X^{A_{1} \dots A_{k}}_{\qquad B_{1} \dots B_{l}} = & +
R^{A_{1}}_{\,\,\, PMN} X^{P A_{2} \dots A_{k}}_{\qquad B_{1} \dots B_{l}} + ... \\
& - R^{P}_{B_{1}MN} X^{A_{1} \dots A_{k}}_{ \qquad P B_{2} \dots B_{l}} - ... 
\label{Parentesi di Lie derivate}
\end{aligned}
\end{equation}

The following systematic procedure can be applied to find a basis of terms entering the anomaly for each separate sector $N_n$:
\begin{enumerate}[a)]
\item Find all the possible subsectors with $N_R$ going  from 0 to the value (\ref{Teorema sottosettori}).
\item Start from $N_R=0$; use repeatedly eq.(\ref{Parentesi di Lie derivate}) to move terms of this subsector to subsectors with higher $N_R$. 
\item Within the remaining terms, apply the relations $I-III$ to restrict the numbers of possible terms in the anomaly basis.
\item Increase $N_R$ by one and repeat the points $b)$ and $c)$.
\item The final basis will be the union of the basis obtained for each subsectors.
\end{enumerate}

\subsection{The sector $ N_n =1$}
We apply the procedure considering the simplest non trivial sector, i.e. $ N_{n}=1 \, . $ 
According to (\ref{Teorema sottosettori}), we have three subsectors $N_R=0,1,2$.
 Without taking into account geometrical constraints,
such as eqs.~(\ref{Bianchi},\ref{Tutte le derivate di Lie}),
the number of possible terms forming the basis of the anomaly is $86$.
The procedure above is a systematic way to select the linearly
independent terms; although it is simple in principle, it
 is quite lengthy and  non-trivial in practical uses; 
it selects just $3$ terms out of $86$. 
The result is summarized in table \ref{All-terms}, 
and the basis is given by the bold terms.

\begin{table}[h!]    
\begin{center}    
\begin{tabular}  {|l|l|} \hline Subsector & Possible terms \\
\hline
$  N_R = 0  $ &  $ D_A D_B D_C D^C D^B n^A \, , \qquad D_A D_C D_B D^C D^B n^A \, , \qquad  D_A D_C D^C D_B D^B n^A \, ,  $  \\
$ N_D = 5 $ & $  D_B D_A D_C D^C D^B n^A \, , \qquad D_B D_C D_A D^C D^B n^A \, , \qquad  D_B D_C D^C D_A D^B n^A \, , $  \\
&  $ D_B D_C D^C D^B D_A n^A \, , \qquad  D_C D_A D_B D^C D^B n^A \, , \qquad
 D_C D_A D^C D_B D^B n^A \, , $  \\
&  $  D_C D_B D_A D^C D^B n^A \, , \qquad D_C D_B D^C D_A D^B n^A \, , \qquad D_C D_B D^C D^B D_A n^A \, , $  \\
&  $ D_C D^C D_A D_B D^B n^A \, , \qquad D_C D^C D_B D_A D^B n^A \, , \qquad D_C D^C D_B D^B D_A n^A  \, . $  \\
  \hline
$ N_R=1 $ &   $ R (D_A D_B D^B n^A)  \, , \,\,\,\,\,\,\, \qquad R (D_B  D_A D^B n^A) \, , \qquad \quad \; R (D_B D^B D_A n^A)  \, ,  $  \\  
$ N_D = 3 $ &   $ R_{BC} (D_A D^C D^B n^A)  \, , \qquad R_{AC} (D_B  D^C D^B n^A) \, , \qquad R_{AB} (D_C D^C D^B n^A)  \, , $  \\
&   $ R_{BC} (D^C  D_A D^B n^A) \, , \qquad  R_{AC} (D^C D_B D^B n^A)  \, , \qquad R_{BC} (D^C  D^B D_A n^A) \, ,   $  \\ 
&   $ R_{ABCD} (D^D D^C D^B n^A)  \, , \,\,\,\,  R_{ACBD} (D^D  D^C D^B n^A) \, , \; \; \, R_{ADBC} (D^D D^C D^B n^A)  \, .  $  \\ 
\cline{2-2}
&   $  ( D_A D_B n^B) D^A R  \, , \qquad \,\,\,\,\,\, (D_B  D_A D_A n^B) D^A R \, , \qquad ( D_B D^B n_A) D_A R  \, , $  \\ 
&   $ ( D_B D^B n^A) D^C R_{AC}  \, , \qquad ( D_A D^B n^A) D^C R_{BC} \, , \qquad ( D^B D_A n^A) D^C R_{BC}   \, ,  $  \\ 
&   $ ( D_C D_B n_A) D^A R^{BC} \, , \qquad ( D_C D_B n_A) D^B R_{AC}   \, , \qquad ( D_C D_B n_A) D^C R^{AB} \, , $  \\ 
&   $  ( D_C D_B n_A) D_D R^{ABCD}   \, , \,\,\,\,\, ( D_C D_B n_A) D_D R^{ACBD}  \, , \; \; \, ( D_C D_B n_A) D_D R^{ADBC}   \, .  $  \\ 
\cline{2-2}
&  $  ( D_A n^A) D_B D^B R   \, , \qquad \,\,\,\,\,\, ( D_B n_A) D^A D^B R   \, , \qquad \quad \; ( D_B n_A) D_B D^A R   \, , $  \\
&  $  ( D_B n^A) D_A D_C R^{BC}   \, , \qquad ( D^B n_A) D_B D_C R^{AC}   \, , \qquad  ( D_B n^A) D_C D_A R^{BC}   \, , $ \\
&  $  ( D^B n_A) D_C D_B R^{AC}   \, , \qquad ( D_A n^A) D_C D_B R^{BC}   \, , \qquad ( D_B n_A) D_C D^C R^{AB}  $   \\
&  $  ( D_B n_A) D_C D_D R^{ABCD}   \, , \,\,\,\,\, ( D_B n_A) D_C D_D R^{ACBD}  \, , \; \; \,  ( D_B n_A) D_C D_D R^{ADBC} \, .  $ \\
\cline{2-2}
&  $  n^A (D_A D_B D^B R )  \, , \qquad \,\,\,\,\,\, n_A (D_B D^A D^B R )  \, , \qquad \quad \; n_A ( D_B D^B D^A R )  \, , $  \\
&  $  n^A (D_A D_C D_B R^{BC} )  \, , \qquad  n_A (D_B D_C D^C R^{AB} )  \, , \qquad  n^A (D_C D_A D_B R^{BC} )  \, ,  $  \\
&  $   n^A (D_C D_B D_A R^{BC} )  \, , \qquad  n_A (D_C D_B D^C R^{AB} )  \, , \qquad  n_A (D_C D^C D_B R^{AB} )  \, ,  $  \\
&  $  n^A (D_B D_D D_C R^{ABCD} )  \, , \,\,\,\,\,\,  n^A (D_B D_D D_C R^{ACBD} )  \, , \; \; \, n^A (D_B D_D D_C R^{ADBC} )\, . $  \\
 \hline
 $N_R =2$ & $ (D_A n^A) R^2  \, , \qquad \qquad \qquad (D_B n_A) R R^{AB} \, , \qquad \qquad    (D_A n^A) R_{BC} R^{BC}  \, , $ \\
 $ N_D = 1 $ & $ (D_B n_A) R^{A}_{\,\,\, C} R^{BC} \, , \qquad \quad \; \, (D_B n_A) R_{CD} R^{ACBD}  \, , $ \\
 & $  (D_A n^A) R^{BCDE} R^{BCDE}  \, , \; \; \, (D_A n^A) R^{BCDE} R^{BDCE}   \, ,  $ \\
 & $ (D_B n_A) R^{BCDE} R^{A}_{\,\,\, CDE}  \, , \quad (D_B n_A) R^{BDCE} R^{A}_{\,\,\, CDE}  \, . $  \\
 \cline{2-2}
 & $ n^A R (D_A R)  \, , \qquad \qquad \qquad \pmb{{ n^A R_{AB} (D^B R)}} \, , \qquad \qquad n_A R (D_B R^{AB})  \, , $ \\
 & $ \pmb{{n_A R_{BC} (D^C R^{AB})}} \, , \qquad \quad \; \; n_A R^{AB} (D^C R_{BC})  \, , \qquad \; \; \, n_A R_{BC} (D^A R^{BC}) \, , $ \\
 & $ n_A R_{BC} (D_D R^{ABCD})  \, , \qquad \; \pmb{{n^A R_{ABCD} (D^D R^{BC})}} \, ,  $ \\
 &  $ n^A R_{ABCD} (D_E R^{BCDE})  \, , \quad n^A R_{ABCD} (D_E R^{BECD}) \, , $ \\
 & $  n_A R_{BCDE} (D^A R^{BCDE})  \, , \quad n_A R_{BCDE} (D^A R^{BDCE})\, , $ \\
  & $ n_A R_{BCDE} (D^C R^{ABDE})  \, , \quad n_A R_{BCDE} (D^E R^{ABCD})\, . $ \\
\hline
\end{tabular}   
\caption{\footnotesize List of all the $86$
scalars with Weyl weight -4 built with 1 vector $ n_A . $  The bold terms compose the final basis for the anomaly.}
\label{All-terms} 
\end{center}
\end{table}

The result is the basis:
\begin{eqnarray}
&  B_1 = (n^A R_{AB}) D^B R \nonumber \\
& B_2 = (D_C R_{AB} ) n^A R^{BC} \\
& B_3= (n^D R_{ABCD} ) D^B R^{AC}  \, , \nonumber  
\end{eqnarray}
and the anomaly:
\beq
\mathcal{A} \supset \int d^3 x \sqrt{g} \s \sum_{k=1}^3  f_k B_k \, .
\label{Aeffe}
\eeq

\subsection{Cohomological problem of the $N_n=1$ sector}

It is possible to show that none of the three terms in the basis can be written as a Weyl variation of any 
combination of the remaining ones. Consequently, there are not counterterms (i.e. scheme dependent terms) in the anomaly.

We can now study the cohomological problem by finding the commutator of two Weyl variations
\begin{equation}
 \Delta_{\sigma_{1}\sigma_{2}}^{\mathrm{WZ}} \mathcal{B}_{k} =  \delta_{\sigma_{1}(x)} 
\int d^{3} x \, \sqrt{g} \, B_{k} \sigma_{2} (y)   
-  \delta_{\sigma_{2}(x)} 
\int d^{3} x \, \sqrt{g} \, B_{k} \sigma_{1} (y)   \, ,
\label{liquirizia}
\end{equation}
for all the terms in the basis, $k=1,2,3.$

Using integration by parts, eq.~(\ref{liquirizia}) can be written as a linear 
combination of $9$ independent expressions $  W_{j} $:
\begin{equation}
\Delta_{\sigma_{1}\sigma_{2}}^{\mathrm{WZ}} \mathcal{B}_{k} =  \int d^{3}x \, \sqrt{g} \, 
\left(  \sum_{k=1}^{9} M^{kj} W_{j}  \right)  \,  ,
\end{equation}
The null space of the matrix $ M^{kj} $ corresponds to the consistent combination entering the anomaly.

The nine independent expressions are:
\begin{eqnarray}
& W_1=\commuA n_B D^AD^B R \, , \qquad
& W_2=\commuA R n_B R^{AB}   \, , \qquad \nonumber  \\
& W_3=\commuA n_B R^{AC} R_{BC}  \, , \qquad & W_4=\commuA R_{BC} n_D R^{ABCD} \, , \nonumber \\
&  W_5=\commuA  (D_B n_C) D^C R^{AB}  \, , \qquad 
& W_6=(D_A \s_{[1} D^2 \s_{2]}) n_B R^{AB} \, ,   \\
& W_7= (D_A \s_{[1} D_B \s_{2]}) n_C D^A R^{BC}  \, , \qquad
& W_8 = (D_A \s_{[1} D_B \s_{2]}) (D_C n^B) R^{AC} \, ,  \nonumber  \\
& W_9 = (D_A \s_{[1} D_B D_C \s_{2]}) n_D R^{ABCD} \, . \nonumber
\end{eqnarray}
The transpose of the matrix $ M^{kj} $ is:
\begin{equation}
(M^t)^{mk}=\left(
\begin{array}{cccccccccccc}
 2 & -1 & \frac{3}{2}  \\
 -2 & 1 & 0  \\
 0 & 0 & -3  \\
 0 & 0 & 3  \\
 0 & 0 & -3  \\
 6 & 1 & 1  \\
 0 & -2 & -1  \\
 0 & -4 & 4 \\
 0 & 0 & -2 \\
\end{array}
\right) \, .
\end{equation}
The null space of this matrix has dimension $0$ and so the coefficients $f_k$
in eq.~(\ref{Aeffe}) must vanish.
There is not  any consistent term which satisfies the Wess-Zumino conditions in the $N_n=1$ sector.

\subsection{Sectors with higher $N_n$}

The problem of the higher $N_n$  sectors is obviously the proliferation of terms in the basis.
Although the procedure  is clear in principle, the problem is intractable on the practical side
unless faced with suitable computer software. This goes beyond the purpose of the present paper.
In principle, the possibility that other type A anomalies exists in higher $N_n$ sectors cannot be discarded. 
What is for sure is that an infinite set of type B anomalies can be found, suitably  combining Weyl tensors with $n$ vectors \cite{Jensen:2014hqa}.
For instance, the combination
\begin{equation}
W_{MNPQ} W^{MNPS} W^{Q}_{\,\,\, ASB} n^A n^B  
\end{equation}
is non-vanishing, belongs to the sector $ N_n =2 $,  has the correct Weyl weight and has vanishing Weyl variation. Similar combinations can be built for any even value of $ N_n $, 
 because by dimensional analysis one Weyl tensor balances 2 $ n_A $ vectors, and the number of type B anomalies is infinite.

\section{Conclusions}

In this paper we initiated the study of the local renormalization
group formalism in the case of a non-relativistic theory
with boost invariance  in $d=2$ spatial dimensions
and nearby a fixed point with dynamical exponent $z=2$. 

If we do not impose the Frobenius
condition, the trace anomaly and the local RG contain an infinite
number of terms, organized in an infinite number of sectors
with decoupled equations. In each sector there is only a finite number
of terms; in particular, in the simplest sector $N_n=0$
the WZ consistency conditions are formally the same 
as in the relativistic case in $4$ space-time dimensions.
Moreover we investigated the structure of the trace anomaly
in the sector $N_n=1$ and we found 
that it is vanishing at the conformal fixed point.

If the Frobenius condition is imposed, the structure of the 
trace anomaly is much simpler; only a finite number of terms
are allowed by dimensional analysis. Moreover, all the quantities
that are constrained by the local RG equation are scheme-dependent at
the fixed point. However, it could be that they still give
useful constraints on the RG flow, similarly to the 
$a'$ conjecture in the relativistic case 
\cite{Anselmi:1999xk,Anselmi:2002fk}.
Indeed, we have found four scheme-dependent quantities $\tilde{A}_{1} \dots \tilde{A}_{4}$
whose evolution is described by an equation
which is similar to the one for $a'$ 
in the relativistic case. An analogous study was performed
 in \cite{Pal:2016rpz} in the case of non-relativistic
theories without boost invariance.

Several directions deserve further investigation:
\begin{itemize}
\item
The proof of a conjectured $a$-theorem in the $N_n=0$ sector
of the anomaly without Frobenius condition
is reduced to the positivity of a quadratic form in the coupling
space $\chi^g_{ij}$.
In spite of the formal similarities with the relativistic case,
the proof does not seem to be trivial.
It would be interesting to check this issue in examples
and eventually to search  for a proof.
\item The anomaly sectors without Frobenius condition
may contain other type $A$ anomalies. Unfortunately
the number of terms quickly proliferates as we increase $N_n$;
a more efficient strategy is needed for a systematic analysis.  
\item We studied in detail just the $z=2$, $d=2$ case.
The DLCQ reduction still works for generic $z$,
provided that one assigns  different Weyl weights to different
components of extra dimensional tensors, see Appendix A of
\cite{Auzzi:2015fgg}. Anomalies can arise only when $d+z$
is an even number; the case that we considered is one of the simplest
and most commonly used in condensed matter applications.
It would be interesting to study other combinations $(d,z)$. 
\item
The scheme-dependent quantities $\tilde{A}_{1} \dots \tilde{A}_{4}$
are constrained by the local 
RG group equation in the Frobenius case.
The monotonicity of these quantities is another open question.
\item In the conjecture proposed in 
\cite{Anselmi:1999xk,Anselmi:2002fk}, the minimum of the 
quantity $\Delta a'$ along all the possible RG flows 
is related to the difference $\Delta a$, which is a
independent on the RG trajectory.
It would interesting to find scheme-independent quantities
related to  $\tilde{A}_{1} \dots \tilde{A}_{4}$ in a similar way.
\end{itemize}

\section*{Acknowledgments}

We are grateful to Diego Redigolo for a useful discussion.

\section*{Appendix}
\addtocontents{toc}{\protect\setcounter{tocdepth}{1}}
\appendix

\section{Local counterterms in the Frobenius case}

\label{controtermini}

Here we will consider  the
following local counterterms in the vacuum functional 
$W$:
\beq
W \rightarrow W+ \int  \sqrt{g} \,  d^3 x  \left[  K_1 R^2 + K_2 (\Omega-2 \chi)^2
+K_3 \chi^2 + K_4 \Omega^2 + K_5 \chi R
+K_6 J^2 \right] \, .
\eeq
We use the Weyl variations in table 
\ref{variazioni}.  

\begin{table}[h]     
\begin{center}    
\begin{tabular}  {|l|l|} \hline Term & Weyl variation \\
\hline
$R$ & $-2 \s R - 6 D^2 \s$  \\
$\chi$ & $-2 \s \chi-\frac{1}{2} w^A \p_A \s$  \\
$\Omega$ & $-2 \s \Omega -w^A \p_A \s +D^2 \s$  \\
\hline
$w_A$ & $ -4 D_A \s$\\
$w^A$ & $-2 \s w^A -4 D^A \s$ \\
\hline
$R_{M N}$ & $-2 D_M D_N \s - G_{M N} D^2 \s $ \\ 
$ \Omega_{CB}$ & $D_B D_C \s-\frac{1}{4} G_{BC} w^K D_K \s$ \\
\hline
$R^2$ & $-4 \s R^2 -12 R D^2 \s $ \\
$(\Omega-2 \chi)^2$ & $-4 \s (\Omega-2 \chi)^2+2 (\Omega-2 \chi) D^2 \s$ \\
$ \chi^2 $ & $ -4 \s \chi^2 -\chi w^A D_A \s $ \\
$ \Omega^2 $ & $ -4 \s \Omega^2 -2 \Omega w^A D_A \s +2 \Omega D^2 \s$ \\
$\chi \Omega$ & $-4 \s \chi \Omega -\le \chi+\frac{1}{2} \Omega \ri w^A D_A \s + \chi D^2 \s $\\
$\chi R$ &$ -4 \s \chi R -\frac{1}{2} R w^A D_A \s -6 \chi D^2 \s$ \\
$\Omega R$ & $-4 \s \Omega R -R w^A D_A \s +(R-6 \Omega) D^2 \s$ \\
$\Omega_{AB} \Omega^{AB}$ & $-4 \s \Omega_{AB} \Omega^{AB}
+2 \Omega_{AB} D^A D^B \s -\frac{1}{2} \Omega w^A \p_A \s$ \\
$\Omega_{AB} w^A w^B$ & $-4\s \Omega_{AB} w^A w^B+
 w^A w^B D_A  D_B \s -4 \chi w^A D_A \s -8 \Omega_{AB} w^A D^B \s $\\
  $w^A D_A R$ & $-4 \s w^A D_A R -4 D^A \s D_A R -2 R w^A D_A \s 
 -6 w^A D_A D^2 \s$ \\
\hline
\end{tabular}   
\caption{\footnotesize Weyl variation of several terms }
\label{variazioni}  
\end{center}
\end{table}

The variation of each term induces a shift in the anomaly coefficients
in eq.~(\ref{LRG-frobe}). The local RG equations in eq.~(\ref{etaeta})
and in appendix \ref{full-list} are  invariant under these shifts;
this provides several non-trivial cross-checks of our calculations.
The list of the shifts induced by each term is:
\begin{enumerate}   
\item Counterterm $K_1$:
\beq
\Delta (K_1 R^2)= \s \b^i \p_i K_1 R^2 -12 K_1 (D^2 \s) R  \, , 
\eeq
\bea
\delta \E^1 &=& \b^i \p_i K_1 \, , \qquad
\delta d^3_i = 24 \p_i K_1 \, , \qquad
\delta \eta^{10}=-12 K_1 \, , \nl
\delta c^1_i &=& 12 \p_i K_1 \, , \qquad
\delta e^1_{ij} =12 \p_{ij} K_1 \, .
\eea
\item Countertem $K_2$:
\beq
\Delta (K_2  (\Omega-2 \chi)^2) =
\s \b^i \p_i K_2  (\Omega-2 \chi)^2 
+ 2 K_2 (D^2 \s) (\Omega-2 \chi) \, ,
\eeq
\bea
& & \v \E^2 = 4 \li(K_2) \, ,\qquad
\v \E^3= \li(K_2) \, ,\qquad
\v \E^4=-4 \li(K_2) \, ,\qquad
\v \E^{11}=-4 K_2 \, ,
\nl
&& \v \E^{12} = 2 K_2 \, , \qquad
\v c_i^2 =4 \p_i K_2 \, , \qquad
\v c_i^3 = -2 \p_i K_2 \, , \qquad
\v d_i^4 =8 \p_i K_2 \, ,
\nl
&&
\v d_i^5 = -4 \p_i K_2 \, , \qquad
\v e_{ij}^2 = 4 \p_{ij} K_2 \, , \qquad
\v e_{ij}^3=- 2 \p_{ij} K_2 \, .
\eea
\item Counterterm $K_3$,
modulo integration by parts:
\bea
\Delta (K_3  \chi^2)&=&\s
\b^i \p_i K_3  \chi^2 - K_3 \chi w^A D_A \s   \nl
&=&
\s \b^i \p_i K_3  \chi^2 
+ \s K_3  \le 12 \chi^2 -4 \chi \Omega
-\frac12 \Omega_{AB} w^A w^B \ri \nl
  && +\s \p_i K_3 \, \chi w^A D_A g^i \, ,
\eea
\beq
\v \E^2= \li(K_3)+12 K_3 \, , \qquad \v \E^4 = -4 K_3 \, , \qquad
\v \E^8 = -\frac12 K_3 \, , \qquad \v c_i^5=\p_i K_3 \, .
\eeq
\item Counteterm $K_4$,
modulo integ by parts:
\bea
\Delta (K_4 \Omega^2) &=& \s \li(K_4) \Omega^2 +2 K_4 \Omega D^2 \s -2 K_4 \Omega w^A D_A \s 
\nl
&=& \s (\li(K_4) -8K_4)\Omega^2 +2 K_4 \Omega D^2 \s
+ \s 2 \p_i K_4 \Omega w^A D_A g^i \nl
&&+\s K_4 \le -3 \Omega_{AB} w^A w^B 
+ 8 \Omega_{AB}^2 -4 D^2 \chi +4 \chi R 
+24 \chi \Omega\ri \, ,
\eea
\bea
&& \v \E^3=\li(K_4) -8K_4 \, , \qquad \v \E^4=24 K_4 \, , \qquad
\v \E^5=4K_4 \, , \qquad \v \E^7 =8 K_4 \, , 
\nl
&& \v \E^8=-3 K_4\, , \qquad \v \E^{11}=-4 K_4 \, , \qquad \v \E^{12}=2 K_4 \, ,
\qquad \v c_i^6 = 2 \p_i K_4 \, ,
\nl
&& \v c_i^3=-2 \p_i K_4 \, , \qquad \v d_i^5 = -4 \p_i K_4 \, ,\qquad
\v e_{ij}^3=-2 \p_{ij} K_4 \, .
\eea
\item Counterterm $K_5$:
\bea
\Delta (K_5 \chi R) &=& \s \li(K_5) \chi R -\frac{K_5}{2} R w^A D_A \s -6 K_5 \chi D^2 \s
\nl
&=& \s \li(K_5)  R \chi 
+K_5 \le 2 \s R \chi 
-2 \s  R \Omega-6  \chi D^2 \s 
+ \frac{1}{2} \s w^A D_A R \ri  \nl
&& +\frac{\s}{2} \p_i K_5 R w^A D_A g^i \,
\eea
\bea
&&\v \E^5=\li(K_5) +2 K_5 \, , \qquad \v \E^6=-2 K_5 \, , \qquad \v \E^9= \frac12 K_5 \, , \qquad
\v \E^{11}=-6 K_5 \, ,
\nl
&& \v c_i^2=6 \p_i K_5 \, , \qquad
\v c_i^4 = \frac12 \p_i K_5 \, , \qquad
 \v d_i^4=12 \p_i K_5 \, , \qquad
\v e^2_{ij}= 6 \p_{ij} K_5 \, .
\eea
\item
Counterterm $K_6$:
\beq
\Delta(F J^2)=\Delta( F (\Omega-2 \chi +R/6)^2)=
\li(F) (\Omega-2 \chi +R/6)^2 \, ,
\eeq
\bea
&& \v  \E^1=\frac{1}{36} \li(F) \, , \qquad
\v \E^2=4 \li(F) \, , \qquad
\v  \E^3= \li(F) \, , \nl
&& \v \E^4=-4 \li(F) \, , \qquad
\v \E^5=-\frac23 \li(F) \, , \qquad
\v \E^6=\frac13 \li(F) \, .
\eea
\end{enumerate}

\section{Consistency conditions with Frobenius conditions}
\label{full-list}

The WZ consistency conditions 
generate the algebraic constraint in eq.~(\ref{etaeta})
and the following system of differential equations:
 \beq
  \d_i(-2 \E^{12} +12 \E^{10})    
  +12 c_i^1 
  -2 \li(d_i^6)
  -4 e_{ij}^6 \beta^j 
  -2 c_i^3 = 0 
 \, , 
  \label{bo4-bis} 
  \eeq
 \bea
  \label{sempli}
 \d_i \le
48  \E^1 +\frac{2}{3} \E^2 -4 \E^3 
-\b^k \le 4 c_k^1  +\frac{2}{3} c_k^3 +\frac{2}{3} c_k^5
 -\frac{4}{3} c_k^6 \ri \ri
 -\li(d_i^1  +16 d_i^2   )  & &\nl 
  + (32 e_{ij}^4 +2 e_{ij}^5 ) \beta^j 
 -  8  (c_i^5  +2 c_i^6) & = & 0  \, , 
\eea
 \bea
   \label{bo8-bis}
 \d_i \le
 -72 \E^1  + 2 \E^3
  -4 \E^{11} -8 \E^{12})
  +\p_i (\b^k( 6 c_k^1 + c_k^3 -4 d_k^7) \ri
   - 2 c_i^2 -4 c_i^3 
   & &
  \nl 
 -24 c_i^4  +4c_i^6  +\li(-d_i^1 -4 d_i^7 )
  + (2 e_{ij}^5 
 +8 e_{(ij)}^7 ) \beta^j &=& 0 \, , 
 \eea
\bea
 \label{bo1-bis}
 \p_i (\E^{11}+2 \E^{12} -6 \E^{10})
 - \d_i (   3 \b^k c_k^1  -2 \b^k d_k^7) 
-  6 c_i^1
 - 3 c_l^1 (\d_i \beta^l) 
 + 3(\d_i c_l^1) \beta^l & & \nl 
 + 12 c_i^4 
+ 3  \li(d_i^3)
 -6 e_{ij}^1 \beta^j  &=&0   \, ,
\eea 
\beq
 \b^k (\d_i d_k^7 - \d_k d_i^7)
 - \frac{1}{2} c_i^2 -c_i^3 
 +6 c_i^4  -c_i^6 +2 \b^j e^7_{[ij]}
  =0  \, , 
 \eeq
 \beq
 -12 c_i^1 
 +4 e_{ij}^6 \beta^j +2 c_i^3 +2\beta^l \d_l d_i^6 
 -d_i^1 +6 d_i^3  - d_i^5 - 2 d_i^6 
  - f_{ij}^1 \beta^j - 2 f_{ij}^2 \beta^j =0  \, ,
\eeq
 \bea
 &&   c_i^3  +   c_l^3 (\d_i \beta^l)  
  -2 \p_i (\b^k d_k^7 ) 
 -12 c_i^4  
 +\frac12 \li(-d_i^1  -d_i^5 )
  + ( e_{ij}^5 
 +2 e_{(ij)}^7
 -2  e_{[ij]}^7
 + e_{ij}^3  
 ) \beta^j  = \nl
  &=& \d_i (   3 \b^k c_k^1
  -2 \b^k d_k^7) 
+3 \le  c_l^1 (\d_i \beta^l)
 - (\d_i c_l^1) \beta^l  \ri   
  + \li(d_i^6-3 d_i^3)
  +(  6 e_{ij}^1 +
   2 e_{ij}^6) \beta^j + c_i^3  -12 c_i^4 =
 \nl
 &=&
 -2 c_i^2 
  -3 c_i^3 
+12 c_i^4 
-4 c_i^6 
 +2\p_i( \b^k d_k^7)
  -2  c_l^2 (\d_i \beta^l) 
   - 3  c_l^3 (\d_i \beta^l) 
 -\li(-d_i^4+8 d_i^2 - \frac32 d_i^5 )
 \nl
 && - (2 e_{ij}^2+3 e_{ij}^3 -16 e_{ij}^4  
 +2 e_{(ij)}^7-2 e^7_{[ij]})\beta^j \, ,
 \eea
 
\bea
 &&-6 \d_j \d_i \E^9 +6 \d_i c_j^4 -\d_i c_j^6  -8 e_{ij}^4 +2 e_{ij}^7 - (\d_i e_{lj}^7) \beta^l +  
  e_{lj}^7 (\d_i \beta^l ) - \beta^l \d_l f_{ij}^3 + (f_{lj}^3 + f_{jl}^3) \d_i \beta^l 
\nl
 &&  + 2 k_{lij}^2 \beta^l 
-\frac{1}{2}\d_i\d_j\E^{11}-\d_i\d_j\E^{12}
  - \frac{1}{2} \p_{j} c_i^2 - \p_{j} c_i^3 - \p_{j} \li(d_i^7) + \p_{j} e_{ik}^7 \beta^k= 0 \, ,
\eea
\beq
-d_l^7 (\d_i \d_j \beta^l) - \frac{1}{2} e_{ij}^2 -e_{ij}^3 -\frac{1}{4} e_{ij}^5  + k_{ijl}^2 \beta^l 
+ \frac{1}{2} \p_{j} c_i^2 + \p_{j} c_i^3 + \p_{j} \li(d_i^7) - \p_{j} e_{ik}^7 \beta^k = 0  \, ,
\eeq
\bea
 && \d_i\d_j(6 \E^{10}-\E^{12})
 +6 \d_j c_i^1 - \d_j c_i^3 - (\d_j \beta^l ) (\d_l d_i^6)  -\beta^l (\d_l \d_j d_i^6) - (\d_i \beta^l) (\d_j d_l^6) - d_l^6 \d_i \d_j \beta^l - e_{ij}^5  
 \nl
&& + 2 e_{il}^6 (\d_j \beta^l) +4 e_{ij}^6 -2 (\d_j e_{il}^6) \beta^l -4 e_{ij}^7 
- \beta^l \d_l f_{ji}^1 -f_{li}^1 (\d_j \beta^l) -f_{jl}^1 (\d_i \beta^l) + 2 k_{lji}^1 \beta^l = 0  \, , 
\eea
\beq
12 e_{ij}^1 -2 e_{ji}^3 -e_{ij}^5 -2 \beta^l \d_l f_{ij}^2 - 2 f_{il}^2 (\d_j \beta^l) 
-2 f_{jl}^2 (\d_i \beta^l) -2 k_{ijm}^1 \beta^m - 2   d_l^6   (\d_i \d_j \beta^l) 
+\p_{ij} (12 \E^{10} -2 \E^{12})= 0 
\eeq
\beq
-\d_i d_j^1 +2 f_{ij}^1 + 2 f_{il}^1 (\d_j \beta^l) -4 f_{ij}^3 + 2 T_{lji} \beta^l = 0
\eeq

\bea
 &-& (\d_i d_l^6) (\d_j \d_k \beta^l) - d_l^6  (\d_j \d_k \d_i \beta^l) + 6 \d_i e_{jk}^1  - \d_i e_{jk}^3 - \d_k e_{ij}^5  - f_{il}^1 (\d_j \d_k \beta^l) 
 -(\d_i \beta^l) (\d_l f_{jk}^2)  
 \nl
  &-& \beta^l (\d_l \d_i f_{jk}^2) -2 (\d_i f_{kl}^2) (\d_j \beta^l) -2 f_{kl}^2 (\d_j \d_i \beta^l) + 2 k_{jki}^1  - \d_i (k_{klj}^1 \beta^l) 
 + k_{jkl}^1 (\d_i \beta^l)  -4 k_{jki}^2
 \nl
 & & 
 - \beta^l (\d_l T_{jki}) - T_{jkl} (\d_i \beta^l) -2 T_{lji} (\d_k \beta^l) + 4 q_{lijk} \beta^l + \d_i\d_j\d_k (6\E^{10}-\E^{12})= 0 \, ,  
 \eea


\begin{thebibliography}{1}

\bibitem{Duff:1993wm}
  M.~J.~Duff,
  Class.\ Quant.\ Grav.\  {\bf 11} (1994) 1387
  [hep-th/9308075].
  
  \bibitem{Zamolodchikov:1986gt}
    A.~B.~Zamolodchikov,
    JETP Lett.\  {\bf 43} (1986) 730
     [Pisma Zh.\ Eksp.\ Teor.\ Fiz.\  {\bf 43} (1986) 565].
 
 \bibitem{Cardy:1988cwa}
  J.~L.~Cardy,
  Phys.\ Lett.\ B {\bf 215} (1988) 749.

 \bibitem{Osborn:1989td}
  H.~Osborn,
  Phys.\ Lett.\ B {\bf 222} (1989) 97.
 
 \bibitem{Jack:1990eb}
  I.~Jack and H.~Osborn,
  Nucl.\ Phys.\ B {\bf 343} (1990) 647.
 
 \bibitem{Osborn:1991gm}
  H.~Osborn,
  Nucl.\ Phys.\ B {\bf 363} (1991) 486.
  
  
\bibitem{Komargodski:2011vj}
  Z.~Komargodski and A.~Schwimmer,
  JHEP {\bf 1112} (2011) 099
  [arXiv:1107.3987 [hep-th]].
 
 \bibitem{Komargodski:2011xv}
  Z.~Komargodski,
  JHEP {\bf 1207} (2012) 069
  [arXiv:1112.4538 [hep-th]].
  

\bibitem{Bonora:1983ff}
  L.~Bonora, P.~Cotta-Ramusino and C.~Reina,
  Phys.\ Lett.\ B {\bf 126} (1983) 305.
  
\bibitem{Bonora:1985cq}
  L.~Bonora, P.~Pasti and M.~Bregola,
  Class.\ Quant.\ Grav.\  {\bf 3} (1986) 635.
  doi:10.1088/0264-9381/3/4/018



\bibitem{Fortin:2012hn}
  J.~F.~Fortin, B.~Grinstein and A.~Stergiou,
  JHEP {\bf 1301} (2013) 184
  doi:10.1007/JHEP01(2013)184
  [arXiv:1208.3674 [hep-th]].

\bibitem{Luty:2012ww}
  M.~A.~Luty, J.~Polchinski and R.~Rattazzi,
  JHEP {\bf 1301} (2013) 152
  doi:10.1007/JHEP01(2013)152
  [arXiv:1204.5221 [hep-th]].

\bibitem{Jack:2013sha}
  I.~Jack and H.~Osborn,
  Nucl.\ Phys.\ B {\bf 883} (2014) 425
  doi:10.1016/j.nuclphysb.2014.03.018
  [arXiv:1312.0428 [hep-th]].


\bibitem{Baume:2014rla}
  F.~Baume, B.~Keren-Zur, R.~Rattazzi and L.~Vitale,
  JHEP {\bf 1408} (2014) 152
  doi:10.1007/JHEP08(2014)152
  [arXiv:1401.5983 [hep-th]].

\bibitem{Nakayama:2013wda}
  Y.~Nakayama,
  Nucl.\ Phys.\ B {\bf 879} (2014) 37
  doi:10.1016/j.nuclphysb.2013.12.002
  [arXiv:1307.8048 [hep-th]].


\bibitem{Stergiou:2016uqq}
  A.~Stergiou, D.~Stone and L.~G.~Vitale,
  JHEP {\bf 1608} (2016) 010
  doi:10.1007/JHEP08(2016)010
  [arXiv:1604.01782 [hep-th]].

\bibitem{Auzzi:2015yia}
  R.~Auzzi and B.~Keren-Zur,
  JHEP {\bf 1505} (2015) 150
  doi:10.1007/JHEP05(2015)150
  [arXiv:1502.05962 [hep-th]].


\bibitem{Gomis:2015yaa}
  J.~Gomis, P.~S.~Hsin, Z.~Komargodski, A.~Schwimmer, N.~Seiberg and S.~Theisen,
  JHEP {\bf 1603} (2016) 022
  doi:10.1007/JHEP03(2016)022
  [arXiv:1509.08511 [hep-th]].

\bibitem{Nakayama:2013is}
  Y.~Nakayama,
  Phys.\ Rept.\  {\bf 569} (2015) 1
  doi:10.1016/j.physrep.2014.12.003
  [arXiv:1302.0884 [hep-th]].


\bibitem{Shore:2016xor}
  G.~M.~Shore,
  arXiv:1601.06662 [hep-th].
  
\bibitem{Anselmi:1997am}
  D.~Anselmi, D.~Z.~Freedman, M.~T.~Grisaru and A.~A.~Johansen,
  Nucl.\ Phys.\ B {\bf 526} (1998) 543
  doi:10.1016/S0550-3213(98)00278-8
  [hep-th/9708042].
  

\bibitem{Anselmi:1999xk}
  D.~Anselmi,
  Annals Phys.\  {\bf 276} (1999) 361
  doi:10.1006/aphy.1999.5949
  [hep-th/9903059].

\bibitem{Anselmi:2002fk}
  D.~Anselmi,
  Class.\ Quant.\ Grav.\  {\bf 21} (2004) 29
  doi:10.1088/0264-9381/21/1/003
  [hep-th/0210124].
  
  
\bibitem{Adam:2009gq}
  I.~Adam, I.~V.~Melnikov and S.~Theisen,
  JHEP {\bf 0909} (2009) 130
  [arXiv:0907.2156 [hep-th]].

\bibitem{Baggio:2011ha}
  M.~Baggio, J.~de Boer and K.~Holsheimer,
  JHEP {\bf 1207} (2012) 099
  [arXiv:1112.6416 [hep-th]].

\bibitem{Griffin:2011xs}
  T.~Griffin, P.~Horava and C.~M.~Melby-Thompson,
  JHEP {\bf 1205} (2012) 010
  [arXiv:1112.5660 [hep-th]].


 
\bibitem{Arav:2014goa}
  I.~Arav, S.~Chapman and Y.~Oz,
  JHEP {\bf 1502} (2015) 078
  doi:10.1007/JHEP02(2015)078
  [arXiv:1410.5831 [hep-th]].

\bibitem{Arav:2016xjc}
  I.~Arav, S.~Chapman and Y.~Oz,
  JHEP {\bf 1606} (2016) 158
  doi:10.1007/JHEP06(2016)158
  [arXiv:1601.06795 [hep-th]].

\bibitem{Pal:2016rpz}
  S.~Pal and B.~Grinstein,
  arXiv:1605.02748 [hep-th].
  
\bibitem{Deser:1993yx}
  S.~Deser and A.~Schwimmer,
  Phys.\ Lett.\ B {\bf 309} (1993) 279
  [hep-th/9302047].

\bibitem{Jensen:2014hqa}
  K.~Jensen,
  arXiv:1412.7750 [hep-th].
  

\bibitem{Auzzi:2016lxb}
  R.~Auzzi and G.~Nardelli,
  JHEP {\bf 1607} (2016) 047
  doi:10.1007/JHEP07(2016)047
  [arXiv:1605.08684 [hep-th]].
  
  
\bibitem{Auzzi:2015fgg}
  R.~Auzzi, S.~Baiguera and G.~Nardelli,
  JHEP {\bf 1602} (2016) 003
   Erratum: [JHEP {\bf 1602} (2016) 177]
  doi:10.1007/JHEP02(2016)003, 10.1007/JHEP02(2016)177
  [arXiv:1511.08150 [hep-th]].

\bibitem{Son:2005rv}
  D.~T.~Son and M.~Wingate,
  Annals Phys.\  {\bf 321} (2006) 197
  [cond-mat/0509786].

\bibitem{Hoyos:2011ez}
  C.~Hoyos and D.~T.~Son,
  Phys.\ Rev.\ Lett.\  {\bf 108} (2012) 066805
  [arXiv:1109.2651 [cond-mat.mes-hall]].

\bibitem{Son:2013rqa}
  D.~T.~Son,
  arXiv:1306.0638 [cond-mat.mes-hall].

\bibitem{Geracie:2014nka}
  M.~Geracie, D.~T.~Son, C.~Wu and S.~F.~Wu,
  Phys.\ Rev.\ D {\bf 91} (2015) 045030
  [arXiv:1407.1252 [cond-mat.mes-hall]].

\bibitem{Jensen:2014aia}
  K.~Jensen,
  arXiv:1408.6855 [hep-th].
  
\bibitem{Hartong:2014pma}
  J.~Hartong, E.~Kiritsis and N.~A.~Obers,
  Phys.\ Rev.\ D {\bf 92} (2015) 066003
  doi:10.1103/PhysRevD.92.066003
  [arXiv:1409.1522 [hep-th]].
  
\bibitem{Hartong:2014oma}
  J.~Hartong, E.~Kiritsis and N.~A.~Obers,
  Phys.\ Lett.\ B {\bf 746} (2015) 318
  doi:10.1016/j.physletb.2015.05.010
  [arXiv:1409.1519 [hep-th]].
  
\bibitem{Hartong:2015wxa}
  J.~Hartong, E.~Kiritsis and N.~A.~Obers,
  JHEP {\bf 1508} (2015) 006
  doi:10.1007/JHEP08(2015)006
  [arXiv:1502.00228 [hep-th]].
  


\bibitem{Duval:1984cj}
  C.~Duval, G.~Burdet, H.~P.~Kunzle and M.~Perrin,
  Phys.\ Rev.\ D {\bf 31} (1985) 1841.


\end{thebibliography}
\end{document}